\documentstyle[12pt]{article}
\input epsf

\newcommand{\frat}[2]{\frac{\textstyle #1}{\textstyle #2}}
\newcommand{\vf}[1]{\mbox{\boldmath $#1$}}
\newcommand{\dmn}[2]{\mbox{$#1\!\cdot\! 10^{#2}\,$}}

\newcommand{\nomer}[1]{\mbox{$\cal N$\hspace{-.5ex}\raisebox{.3ex}
           {\underline{\tiny 0}$\!$} #1}}

\begin{document}

\begin{center}
{\Large{\bf Potential of Interaction between Non-Abelian Charges in the
Yang-Mills-Higgs Model }}
\end{center}

\begin{center}
{\large S.V. Molodtsov} 
\end{center}

\begin{center}
{\normalsize {\sl State Research Center Institute of Theoretical and
Experimental Physics,}}\\
{\normalsize     {\it Bol'shaya Cheremushkinskaya ul. 25, Moscow,
117259, Russia}}
\end{center}

\begin{abstract}
The problem of two nonrelativistic chromoelectric and chromomagnetic
charges in a Higgs vacuum is considered in classical field theory. An
approximation of interaction potential is constructed on the basis of a
numerical solution to the equations of gluostatics. The concept of a
non-Abelian Abrikosov vortex is discussed. It is shown that the results
of Ginzburg-Landau theory for the tension of a string between magnetic
charges can be directly extended to the non-Abelian case.
\end{abstract}

\section*{ \bf Introduction}
At present, the mechanism of dual superconductivity is one of the most
appealing concepts that are invoked in attempts at explaining quark 
confinement \cite{Sim}.
It is assumed that the physical vacuum hinders the penetration of gluon
fields to large distances from the sources of charge and compresses
lines of force in the same manner as tube does. This pattern is well
known in the Abelian theory of superconductivity for magnetic charges.
The mean-field approximation as exemplified by Ginzburg-Landau theory
is quite sufficient for obtaining a quantitative description of arising
physical situation--and in particular, for evaluating the tension coefficient
\cite{Ball}.
Since the microscopic theory of confinement has yet to be developed,
the above furnishes sufficient motivation for performing a meanfield
analysis of the problem of non-Abelian charges in a vacuum that
possesses the properties of a gluon superconductor. In this formulation
of the problem, Yang-Mills theory is supplemented with Higgs fields,
and the response of a medium to the presence of non-Abelian charges 
( at the macroscopic level) is described by two phenomenological
constants in a phenomenological Lagrangian. An attempt of this kind has
recently been made in \cite{Possib}, 
where the approximation of gluostatics developed earlier for the case
of trivial vacuum \cite{Kh}--\cite{Gol}
was used to describe nonrelativistic heavy particles carrying
non-Abelian charges. It was shown that the equations of gluostatics
with Higgs fields are consistent and that these equations describe
Yang-Mills fields generated by the charges themselves and the response
of the vacuum to the presence of these charges. Some general properties
of solutions to these equations and the possible scenarios of charge
confinement have also been discussed.

This article reports on some results obtained from a numerical analysis
of the equations of gluostatics involving Higgs fields. The results of
numerical calculations are used to construct a simple approximation for
the potential interaction between non-Abelian charges.
By means of a direct substitution, it is shown that there are
vortex-type excitations in the Yang-MIlls-Higgs system and that these
excitations are identical to Abrikosov vortices. Owing to this, it turns
out that the numerical results for the tension coefficient that were
obtained in \cite{Ball} for a funnel-shaped potential of interaction
between magnetic charges in a superconductor directly apply to
chromomagnetic charges in a nontrivial vacuum.

\section{ Chromoelectric charges}

In the model in question, the Lagrangian density has the form
\begin{equation}
\label{eq1}
L=-\frat{1}{4} \widetilde G^{\mu\nu} \widetilde G_{\mu\nu} -
\widetilde j^\mu \widetilde A_\mu
-\frat{1}{2} D^{\mu} \widetilde \chi D_{\mu}\widetilde \chi - 
\frat{\lambda}{4} \{\widetilde \chi^2-F^2 \}^2,
\end{equation}
where $ \widetilde A_\mu$ and $\widetilde \chi$ 
are the triplets of, respectively, Yang-MIlls and Higgs fields 
(~hereafter, the analysis is restricted to the case of the $SU(2)$ group);
$\widetilde j^\mu$ is the density of the external-source currents;
$\widetilde G_{\mu\nu}= \partial_\mu \widetilde A_\nu 
-\partial_\nu \widetilde A_\mu + g
\widetilde A_\mu \times \widetilde A_\nu$ 
is the strength tensor of the gluon field ( the cross denotes vector
product in isotopic space); and the covariant derivative is defined as 
$ D^\mu \widetilde \varphi =  \partial^\mu 
\widetilde \varphi  + g \widetilde A^\mu \times \widetilde \varphi.$

The equations for the Yang-Mills and Higgs fields are obviously given by
\begin{eqnarray}
\label{eq2}
&& D^{\mu}\widetilde G_{\mu\nu}+
g  D_{\nu} \widetilde \chi \times \widetilde \chi = \widetilde j_\nu,
\nonumber\\ [-.2cm]
\\[-.25cm]
&& D^{\mu} D_{\mu}\widetilde \chi  - \lambda \{\widetilde \chi^2-F^2  \}
  \widetilde \chi=0. \nonumber
\end{eqnarray}

In the gluostatic approximation, we retain only the zeroth component of
the $4$-current. We have
$$\widetilde j^0\!=\!\widetilde\rho\!=
\!\widetilde P_1\, \delta({\vf x}-{\vf x}_1) 
+ \widetilde P_2\, \delta({\vf x}-{\vf x}_2),$$
where $P_i, {\vf x}_i,\;i=1,2$ are the vector-charges and coordinates
of the particles involved. Further, the vector-charges of the particles
represent a convenient basis---$\widetilde P_1, \widetilde P_2,\;
\widetilde P_3\!=\!\widetilde P_1\! \times\! \widetilde P_2$
---in which we expand solutions to equations (\ref{eq2}).
The zeroth component $\widetilde A_0\!=\!\widetilde \varphi$
of the gluon field then appears as a linear combination of the
vector-charges of the particles,
\begin{equation}
\label{eq3}
 \widetilde \varphi({\vf x},t)=\varphi_1({\vf x}) 
\widetilde P_1(t)+\varphi_2 ({\vf x})\widetilde P_2(t),
\end{equation}
while the vector field is proportional to the third component of the
three basis vectors:
\begin{equation}
\label{eq4}
\widetilde {\vf A}({\vf x},t)={\vf a}({\vf x}) \widetilde P_3(t).
\end{equation}
The potentials and the vector fields are functions of the spatial
coordinate and particle coordinates:
$\varphi_i({\vf x}|{\vf x}_1,{\vf x}_2), \;
 {\vf a}({\vf x}|{\vf x}_1,{\vf x}_2).$

By virtue of the condition
$$ \partial^{\mu}  \widetilde j_{\mu}+ 
g \widetilde A^\mu \times \widetilde j_{\mu} =0,$$
which ensures the consistency of equations (\ref{eq2}),
the basis generally proves to be rotating in isotopic space in the
direction of the vector
$\widetilde \Omega\! = \stackrel{*}{\varphi}_1 \!\!\widetilde P_1 +
\stackrel{*}{\varphi}_2\!\! \widetilde P_2,$ 
with frequency $|\widetilde \Omega|,$ whereas the functions 
$\stackrel{*}{\varphi_1}=\!\!\varphi_1({\vf x}_2)~~,~~{\mbox{ and}}~~
\stackrel{*}{\varphi_2}=\!\!\varphi_2({\vf x}_1)$ 
determine the values of the potentials and the points where the charges
reside.

Nontrivial solutions to the system of equations
(\ref{eq2}) were studied in \cite{Possib} 
for the Higgs field represented as a linear combination of the
vector-charges of the particles
\begin{equation}
\label{eq5}
 \widetilde \chi({\vf x},t)=\chi_1({\vf x}) 
\widetilde P_1(t)+\chi_2 ({\vf x})\widetilde P_2(t). 
\end{equation}
Following factorization of the contributions of the resulting functions
${\vf a}~,$\\
$\varphi^{\mbox{\tiny T}}=\| \varphi_1~,~~\varphi_2\|~,~~
{\mbox {and}}~~
\chi^{\mbox{\tiny T}}=\| \chi_1~,~~\chi_2~\|,$
the system of equations of gluostatics assumes the form
\begin{eqnarray}
\label{eq6}
&&{\vf D}\!{\vf D}\Phi
+g^2\{ \Phi J\chi \}C\chi
={ \delta}, \nonumber\\
&&{\vf D}\!{\vf D} \chi
+g^2\{ \Phi J\chi \}C\Phi=\lambda \{ \chi JC\chi -F^2\}\chi, 
\nonumber\\ [-.2cm]
\\[-.25cm] 
&& \nabla\!\!\times\!\!\nabla\!\!\times\!{\vf a}-g{\vf j}_{\varphi}+
g{\vf j}_{\chi}=0, \nonumber\\
&&{\vf j}_{\varphi}=\Phi J\!{\vf D} \Phi,
 \;\; {\vf j}_{\chi}= \chi J\!{\vf D}\chi.\nonumber
\end{eqnarray}
where the column $\Phi$ is the difference of the columns $\varphi$
and $\stackrel{*}{\varphi},\;
\stackrel{*}{\varphi}^{\mbox{\tiny T}}=\|\stackrel{*}{\varphi}_1,
\stackrel{*}{\varphi_2}\|$:
$\Phi=\varphi-\!\stackrel{*}{\varphi},\\
{\vf D}_{kl}=\nabla \delta_{kl}+g{\vf a} C_{kl}\;
(\; k,l=1,2)$ is the covariant derivative,\\
$ \delta^{\mbox{\tiny T}}=\|\delta ({\vf x}-{\vf x}_1), 
\delta ({\vf x}-{\vf x}_2)\|,$
and $ C$ and $ J$ are the $2\!\times\!2$ matrices

\vspace{0.25cm}
\parbox[b]{3.in}{$%{3.6in}
C=
 \left \| \begin{array}{rr}
-(\widetilde P_1 \widetilde P_2) & -(\widetilde P_2 \widetilde P_2) \\
(\widetilde P_1 \widetilde P_1) & (\widetilde P_1 \widetilde P_2)
\end{array} \right \|,$}
\parbox[b]{3.in}{$%{3.6in}
J=
 \left \| \begin{array}{rr}
 0 & 1 \\
-1 & 0
\end{array}  \right \|.$}
%\bigskip

\vspace{0.25cm}

The parentheses in the above expression denote scalar products of
vector-charges in isotopic space, and $\delta$ is the delta function
that describes the charge source of intensity equal to unit.

A more general formulation assumes the use of additional components
that have nonzero projections on other basis vectors as well. In
contrast to electrodynamics, non-Abelian theory involves an extra
degree of freedom associated with the choice of the relative
orientation of the Yang-Mills and Higgs fields in isotopic space. The
resulting system of equations is rather cumbersome, but the
regularities in the changes that its solutions suffer as the result of
this generalization may be traced by considering, instead of 
(\ref{eq5}), the limiting case in which the Higgs field have a nonzero
projection only onto the third vector of the basis: 
\begin{equation}
\label{eq7}
 \widetilde \chi({\vf x},t)=\chi({\vf x})\, \widetilde P_3(t). 
\end{equation} 
Instead of the system of equations (\ref{eq6}), we then have
\begin{eqnarray}
\label{eq8}
&&{\vf D}\!{\vf D}\Phi
-g^2 \chi^2 \Phi= \delta, \nonumber\\
&&\triangle \chi-g^2\widetilde \Phi^2 \chi=\lambda \{ \chi^2 -F^2\}\chi, \\
&& \nabla\!\!\times\!\!\nabla\!\!\times\!{\vf a}-g{\vf j}_{\varphi}=0,
\;{\vf j}_{\varphi}=\Phi J\!{\vf D} \Phi.\nonumber
\end{eqnarray}

Let us supplement this system of equations with boundary conditions.
The delta sources on the righthand sides of (\ref{eq6}) and (\ref{eq8})
are eliminated by isolating the Coulomb term in the solutions for the field
$\varphi$  as
$\varphi_c$
$$\varphi^{'}=\varphi_{c}+\varphi.$$
Experience gained from the calculations and the analysis performed
previously for the equations of gluostatics in the trivial vacuum
reveal that, at large values of the coupling constant $g$, we must
approximate pointlike source more accurately, taking into account
induced charge density \cite{Mat}. However, for the parameter values
considered below, the above-type superposition involving the Coulomb
solution and simple boundary conditions at the points where the charges
are located are quite sufficient for achieving a reasonably high
accuracy of the calculations and a reasonably fast convergence of
iterations.  The quantities $\stackrel{*}{\varphi}$ 
were treated as parameters in one version of the calculations and were
determined in solving numerically the problem with free boundary
conditions of the second kind at the points where the charges are located
$\left.\frat{\partial\varphi_1}
{\partial {\vf n}}\right|_{{\vf x}={\vf x}_2}=0,
\;\left.\frat{\partial\varphi_2}
{\partial {\vf n}}\right|_{{\vf x}={\vf x}_1}=0.$

An advantageous feature of the rotating basis is that the boundary
conditions at spatial infinity ( when the calculation is performed in a
large box) are specified in a very simple form:
$\varphi\,|_{\mbox{\tiny G}}\to 0,\; a_{\rho,z}\,|_{\mbox{\tiny G}}\to 0.$
In this case, the Higgs field must approach the vacuum value
$\widetilde \chi^2=F^2,$ whence it follows, in particular, that, if the
charges are equal in magnitude
($|\widetilde P_1|=|\widetilde P_2|=P$), we have
$\chi_1\,|_{\mbox{\tiny G}}=\chi_2\,|_{\mbox{\tiny G}}=\chi_{as}=
F/(2\,P\,\cos(\theta/2))$, where $\theta$ is the angle between the
vector-charges in isotopic space ( this solution is obviously singular at
$\theta=\pi$). In the case specified by equation (\ref{eq7}), the
boundary condition at infinity has the form  $\chi\,|_{\mbox{\tiny G}}=F.$ 
The formulation of the problem admits the possibility of imposing
additional first- and second-order boundary conditions on the field
$\widetilde\chi$ that are analogous to the above boundary conditions
for the fields  $\varphi$ at the points where the charges reside.

The potential of interaction between the two charges is determined by
integrating over space the symmetrized field-energy density
\begin{eqnarray}
\label{eq9}
&&T={\cal E}_e+{\cal E}_m, \nonumber\\
&&{\cal E}_e=\frat{1}{2}{\vf E}_\Phi JC{\vf E}_\Phi+
\frat{g^2}{2}\widetilde P_3{^2}\{ \Phi J\chi \}^2
+\frat{1}{2}{\vf E}_\chi JC{\vf E}_\chi+
\frat{\lambda}{4} \{\widetilde \chi^2-F^2 \}^2,\\
&&{\cal E}_m=\frat{1}{2}\widetilde P_3{^2}{\vf H}^{2},\nonumber
\end{eqnarray}
where ${\vf E}_\Phi={\vf D}\Phi,\;{\vf E}_\chi={\vf D}\chi,$ and
${\vf H}=\nabla\!\!\times\!{\vf a}.$

Integration by parts makes it possible to recast the expression for the
energy density ${\cal E}_e$ of the gluoelectric field into the form
\begin{eqnarray}
\label{eq9a}
&{\cal E}_e=-\frat{1}{2}\varphi JC\delta-
\frat{g^2}{2}\widetilde P_3{^2}\{ \stackrel{*}\varphi\!\! J\chi \}
\{ \Phi J\chi \}-\frat{g}{2}\widetilde P_3{^2}
\{ \stackrel{*}\varphi\!\! J{\vf E}_\Phi \}{\vf a}+\nonumber\\ [-.2cm]
\\[-.25cm] 
&+\frat{g^2}{2}\widetilde P_3{^2}\{ \Phi J\chi \}^2
+\frat{\lambda}{4} \{F^4\!-\widetilde \chi^4 \},
\nonumber
\end{eqnarray}
which is more convenient for numerical calculations.

Let us first consider the case in which the Higgs field is specified by
equation (\ref{eq5}) and analyze solutions to the system of equations
(\ref{eq6}). The energy of the gluomagnetic field and the contributions
to the energy that come from the five terms in (\ref{eq9a}) are
presented in the table as functions of the distance between the charges
[~after integration over the space of energy densities with our
numerical solution to the system of equations (\ref{eq6})]. 
For the parameters, we choose the following characteristic values:
$g=1,\;P=1,\;\lambda=1,\;F=1,$ and $\theta=\pi-\pi/10$ 
( in connection with the choice of $\theta,$ it should be noted that,
in numerical calculations, the vector charges cannot be taken to be
strictly parallel because $\chi_{as}$ is singular in this configuration).
The calculations were performed for the free boundary conditions at the
points where the charges are located. If the boundary conditions
of the first order are used and if the parameter values are chosen as
above, the minimum of energy can be achieved by varying 
$\stackrel{*}\varphi;$ as a result, we arrive at the same solution. It
should also be noted that the singular contributions that describe
self-interaction were eliminated in the same way as in electrodynamics. 
 
From the table, we can see that the fourth and fifth terms represent
the largest corrections to the first term, but they are of opposite
signs and compensate each other. The second and third terms and the
contribution of the gluomagnetic energy are negligibly small. Following
a natural regularization of divergences, the first term is expressed in
terms of $\stackrel{*}{\varphi}$ and takes the form
\begin{equation}
\label{eq10}
V_{int}=-\frat{\stackrel{*}{\varphi}_1+\stackrel{*}{\varphi}_2}{2}\;
(\widetilde P_1 \widetilde P_2).
\end{equation}
Figure 1 displays the potential $V_{int}$ as a function of the distance
$r$ between the charges for $F=0,1,2,$ and $3$
( $\lambda,\;g,\;P,$ and $\theta$ are specified as above).
For the trivial condensate $F~=~0,$ we naturally arrive at the Coulomb
interaction represented in Fig. 1 by points. As $F$ is increased, the
curves are shifted below, this shift being equidistant at large $r.$ 

In discussing the data presented in the table, we have already noted
that, owing to cancellation of the leading corrections, the interaction
potential is determined primarily only by the values of the fields
$\varphi$ at the points where the charges reside. Thus, the detailed
distribution of the vector and Higgs fields may prove not very
important, and even a rough approximation to the numerical solution
will yield a reasonable result for the interaction potential. Indeed,
let us assume that the vector field is negligibly small in the system
of equations (\ref{eq6}), and let us approximate the higgs field in the
entire space by its asymptotic value. The equations for the fields 
$\varphi$ then become trivial, and the required solution is found
straightforwardly ( the angle $\theta$ drops from final expressions).
As a result, we obtain
\begin{eqnarray}
\label{eq11} 
\varphi_1=-\frat{1+e^{-gF\,|x- x_1|}}{8\pi\,|x-x_1|}
-\frat{1-e^{-gF\,|x-x_2|}}{8\pi\,|x-x_2|},\nonumber\\ [-.2cm]
\\[-.25cm] 
\varphi_2=-\frat{1-e^{-gF\,|x-x_1|}}{8\pi\,|x-x_1|}
-\frat{1+e^{-gF\,|x-x_2|}}{8\pi\,|x-x_2|}.\nonumber
\end{eqnarray}
In the difference $\varphi_1-\varphi_2,$ the Coulomb components cancel
out completely. If this were not the case, the fourth term in
expression would guarantee a linear growth of the potential
( this provides a convenient tool for checking the accuracy of our
calculations)\cite{Possib}. 

The required constants $\stackrel{*}{\varphi}$ are now given by
\begin{equation}
\label{eq12a}
\stackrel{*}{\varphi}_1=\stackrel{*}{\varphi}_2=
-\frat{1+e^{-gF\,|x_1-x_2|}}{8\pi\,|x_1-x_2|}
-\frat{gF}{8\pi}.
\nonumber
\end{equation}
The interaction potential can then be approximated as
\begin{equation}
\label{eq12b}
V_{int}=\frat{(\widetilde P_1 \widetilde P_2)}{8\pi} \,
\left\{\frat{1+e^{-gF r}}{r}+gF\right\}.
\end{equation}
The following comment is in order. Only for $\lambda \gg 1$ 
could we hope that the above crude approximation is reasonable. This is
the limit in which the Higgs fields differ noticeable from their
asymptotic values in the region of dimension $r \le \lambda^{-1/2}$
and make a negligibly small contribution to the total energy integral.
But owing to cancellation of the significant corrections, this
approximation proves valid down to $\lambda \sim 1,$ which is a
surprising result.

Figure 2 shows the interaction potential as a function of the coupling
constant $g$ at a fixed distance between the particles ( $r=5$).
On the curve presented in this figure, we can find the $d$ value
corresponding to the onset of a nonlinear regime ( it is quite
conceivable, however, that this restriction is peculiar to the
computational algorithm that we used and has nothing to do with real
physics). 

It was indicated in \cite{Possib} that the mode of solutions that is
characterized by the suppression of the vector field 
$|{\vf a}|=0,\;{\vf j}_{\chi}={\vf j}_{\Phi}$
must manifest itself with increasing $\lambda.$
Figure 3 shows the energy of the gluomagnetic field as a function of
$\lambda.$ We can see that the vector field in the system is suppressed
as the superconducting properties of the medium are enhanced.

It only remains for us to consider the case described by the system of
equations (\ref{eq8}), which corresponds to the Higgs field
proportional to the third basis vector. It can easy be seen that the
charges are now coupled only by short-range forces because, in this
field configuration, the gluoelectric fields are screened at large
distances, the screening factor being $e^{-g F r}.$ For the same
reason, the generation of gluomagnetic fields is insignificant.
As a result, the charges do not interact at large distances, so that,
instead of (\ref{eq12b}) we approximately have
\begin{equation}
\label{eq14}
V_{int}=\frat{(\widetilde P_1 \widetilde P_2)}{4\pi}\,\frat{e^{-g F r}}{r}.
\end{equation}

From the above results, it follows that solutions of the type
(\ref{eq5}) are energetically favorable in the case of attraction
between the particles [ $(\widetilde P_1 \widetilde P_2)<0$],
whereas the configuration (\ref{eq7}) is preferable in the case of
repulsion. All these solutions are stable in classical field theory.
Solutions are determined (``strengthted'') by the boundary conditions, 
which specify the choice of the transverse component of the Higgs 
field lying in the plane spanned by the vector-charges of the particles. 
An effective potential that is expected to arise in quantum theory 
will contain the above solutions with a factor of continual integration.

A simple comparison of the potential (\ref{eq12b}) with the potential 
used in the model of heavy quarkonia ( see \cite{Vain}), 
$$V(r)=-\frat{\alpha}{r}+\beta r+ V_0,
$$
where $\alpha=0.27,\;\beta=0.25 {\mbox{ GeV}}^2,$ and 
$V_0=-0.76 {\mbox{ GeV}}$ ( $r$ is measured in GeV$^{-1}$),  
leads to the following estimates of the constants:
$$g\cong2.6,\;F\cong1.1,\;\;(\; \frat{g^2}{8\,\pi}\cong0.27,\;
\frat{g^3 F}{8\,\pi}\cong0.76).
$$
At the same time, potential models yield a much lower value for the
coupling constant $g$ \cite{Byc}:
$$g(J/\psi)\cong1.55,\;\;g(\Upsilon)\cong1.42,\;
\alpha_s(J/\psi)\cong0.19,\;\;\alpha_s(\Upsilon)\cong0.16.
$$
Conceivable, this is indicative of the importance of loop corrections.
The relationship between the results presented in this study and those
produced by the potential models of heavy quarkonia will be studied
elsewhere in greater detail.

In this section, we have discussed the case in which the non-Abelian
charges are placed in a strong gluonic superconductor. For the sake of
completeness , we present an approximation for the case in which the
vacuum reduces to a trivial one (~$\lambda \to 0,\; F\to 0$). 
For not overly large coupling constants ( $g^2/(4 \pi)<1$),
the interaction potential that takes into account the contribution of
the gluomagnetic field has the form \cite{Gol}
\begin{eqnarray}
\label{eq15}
V_{int}=\frat{(\widetilde P_1 \widetilde P_2) + 
\gamma \widetilde P^2_3}{4\pi\,r},\;
\gamma=\frat{g^2}{4 \pi}\,
\frat{6-\pi^2/2}{16 \pi}.
\end{eqnarray}

\section{ Chromomagnetic charges} 
It was noted in \cite{Possib} that the system of equations (\ref{eq7}), 
which corresponds to the approximation of gluostatics, can also be used
to describe magnetic charges. We will illustrate this statement by
means of an explicit substitution. In the absence of chromoelectric
charges, we begin by eliminating singular delta-function sources from
the system of equations and seek a particular solution of the form
$\Phi=0$ ( in the case under study, the column $\stackrel{*}\varphi$ 
is considered as a free parameter).

The resulting system of equations,
\begin{eqnarray}
\label{eq16}
&&{\vf D}\!{\vf D} \chi=\lambda \{ \chi JC\chi -F^2\}\chi, 
\nonumber\\ [-.2cm]
\\[-.25cm] 
&& \nabla\!\!\times\!\!\nabla\!\!\times\!{\vf a}+
g{\vf j}_{\chi}=0,\;\; {\vf j}_{\chi}= \chi J\!{\vf D}\chi,\nonumber
\end{eqnarray} 
describes the expulsion of a gluomagnetic field from a Higgs condensate
just in the same way as this occurs in the Abelian theory of
superconductivity. From the symmetry properties of the system of
equations and from the fact that the eigenvalues of the matrix $C$ 
are complex conjugate to each other
(~$\mu_{1,2}= \mp i |\widetilde P_3|$), it follows that a Higgs doublet
can be described by one complexvalued function. The equations of the
theory are then equivalent to the Ginzburg-Landau system of equations.
Explicitly, this can be demonstrated by means of the substitutions
\begin{eqnarray}
\label{eq17}
\chi_1=\left\{\frat{\psi}{1-e^{-i\theta}}+
\frat{i\stackrel{*}\psi}{1+e^{i\theta}}
\right\}\frat{F}{\sqrt{2\,|\widetilde P_3|}},\nonumber\\ [-.2cm]
\\[-.25cm] 
\chi_2=\left\{\frat{\psi}{1-e^{i\theta}}-
\frat{i\stackrel{*}\psi}{1+e^{-i\theta}}
\right\}\frat{F}{\sqrt{2\,|\widetilde P_3|}}.\nonumber
\end{eqnarray} 
Instead of two real-valued functions, transformations (\ref{eq17}) 
define a complex-valued function $\psi$ ( here, $\stackrel{*}\psi$ 
is the complex conjugate of $\psi$). 
In addition, we introduce the scale transformations $\nabla \to g F\;
 \nabla$ and ${\vf a} \to \frat{F}{|\widetilde P_3|}\;{\vf a},$ which
reduce the equations for the function $\psi$ and the vector field ${\vf a}$ 
to the canonical form
\begin{eqnarray}
\label{eq18}
&& (\nabla -i {\vf a})^2 \psi=\kappa \{ |\psi|^2 -1\}\psi, 
\nonumber\\ [-.2cm]
\\[-.25cm] 
&& \nabla\!\!\times\!\!\nabla\!\!\times\!{\vf a}-
\frat{i}{2} \{\psi (\nabla +i {\vf a})\!\stackrel{*}\psi-\stackrel{*}\psi
(\nabla -i {\vf a})\psi\}=0, \nonumber
\end{eqnarray}
where $\kappa=\frat{\lambda}{g^2}.$

It is well known  \cite{Land} that the system of equations (\ref{eq18}) 
has solutions with a quantized magnetic flux.
In the initial variables, these solutions correspond to Abrikosov
vortices with the magnetic flux
$$\oint {\vf a}\, d{\vf l}=Q_n=\frat{2 \pi\, n}{g |\widetilde P_3|},
\;n=\pm 1, \pm2, \ldots,$$
where ${\vf l}$ is a closed contour that circumvents the vortex un the
unperturbed condensate. Formally, the flux of the gluomagnetic field is
then given by
$$\oint \widetilde {\vf A}\, d{\vf l}=\frat{2 \pi\, n}{g}
\frat{\widetilde P_3}{|\widetilde P_3|}.$$

The gluoelectric field identically vanishes
$$\widetilde {\vf E}= \widetilde P_1 {\vf D}\Phi_1+
\widetilde P_2 {\vf D}\Phi_2~\equiv~0~,$$
and so does the gauge-invariant 't Hooft tensor of the field strength;
that is, we have
$$ F_{\mu \nu}= \widehat \chi^a G_{\mu \nu}^a -
\frat{1}{g}\, \widehat \chi^a ( D_\mu \widehat \chi \times D_\nu 
\widehat \chi)^a \equiv 0,$$
where $\widehat \chi=\widetilde \chi/|\widetilde \chi|.$
The energy density is concentrated near the vortex core and is given by
\begin{eqnarray}
\label{eq19}
E=g^2 F^4\; {\cal E},~~~ 
{\cal E}=\frat{1}{2}(\nabla \times {\vf a})^2+
\frat{1}{2} |(\nabla -i {\vf a})\psi|^2+
\frat{\kappa}{4} \{ |\psi|^2-1 \}^2.
\end{eqnarray}
At $\kappa=1/\sqrt{2},$ the integral of energy is quantized and has the
linear density $\int E_n\, d {\vf s}= \pi\, F^2\, n.$
In the Abelian case, solutions with large $|n|$ are stable for
$\kappa<1/\sqrt{2}$ \cite{Bg}. In the case being considered, the
gluomagnetic field does not necessarily disturb a gluonic
superconductor, because there is an additional degree of freedom
associated with the relative orientation of the Yang-Mills and Higgs
fields in isotopic space. Thus, solutions of the Abrikosov-vortex type
are metastable, but they may lie comparatively far from the instability
region. Solutions of this type have already been used in the theory of
electroweak interaction and are referred to as $W$ and $Z$
strings \cite{Vac}.

Unfortunately, magnetic charges separated by large distances can be
described only by solving numerically equations (\ref{eq18}) with the
input singular Dirac potential
\begin{equation}
\label{eq20}
A_{\mbox{\tiny D}}=\frat{g}{4 \pi} 
\frat{{\vf e}_\varphi}{\rho} \left \{ 
\frat{z-d}{[\rho^2+(z-d)^2]^{1/2}}-\frat{z+d}{[\rho^2+(z+d)^2]^{1/2}}
\right\},
\end{equation}
where ${\vf e}_\varphi$ is a unit vector, $\rho$ and $z$ are the radial
coordinates, and $\pm d$ are the points at the $z$ axis where the
magnetic poles reside. The results of such numerical investigations are
presented in \cite{Ball}. among other things, the tension coefficient
for a funnel-like potential was calculated in these studies as a
function of the parameters of the Ginzburg-Landau potential. Formulas 
(\ref{eq17}), which establish relationship between the Abelian and
non-Abelian models, can be used to extend these results to the case of
chromomagnetic charges in a gluonic superconductor. 

\section*{Conclusion}
The problem of two nonrelativistic chromoelectric and chromomagnetic
char\-ges in a Higgs vacuum with the properties of a gluonic
superconductor has been considered within classical field theory. It
has been shown that there is a direct analogy with the Abelian theory
of superconductivity, where magnetic charges are confined, while
electric charges are not ( by charge confinement, we mean here a linear
growth of the energy of particle interaction with distance). It has
been found that Abelian modes do not exhaust all possible states of
chromoelectric charges; for these, there exist states whose energy is
less than the Coulomb ( Yukawa) energy. It is conceivable that the
inclusion of loop corrections may reconcile the results presented in
this study with available data on the potential of heavy quarkonia,
thereby ensuring confinement of charges.

A simple representation is given above for a certain class of solutions
to the Yang-Mills-Higgs equations that describes non-Abelian Abrikosov
vortices. The solutions obtained here can be used in constructing
models of elementary particles, which can then be considered either as
closed vortex lines in the bulk of a Higgs condensate or as droplets of
a dual condensate with a quantized gluoelectric field frozen into them.
In the latter case, the source and outflux of the gluoelectric field
from the droplet surface can be interpreted as a quark-antiquark pair.

Considering the dual pattern, we arrive, for the dual fields (
potentials) and charges, at the scenario in which the fates of
chromoelectric and chromomagnetic charges are interchanged: the former
are confined, while the latter are free.
\\
\\
\noindent{\bf Acknowledgments}. 
The work was supported by russian Foundation for Basic Research 
( project \nomer{96-02-16303-a}).

\newpage
\begin{center}
Gluoelectric and gluomagnetic contributions to the energy
\end{center}
\begin{tabular}{|ccccccc|}
\hline
$r$&$V_e^{{\mbox{\tiny I}}}$&$V_e^{{\mbox{\tiny II}}}$
&$V_e^{{\mbox{\tiny III}}}$&$V_e^{{\mbox{\tiny IV}}}$
&$V_e^{{\mbox{\tiny V}}}$&$V_m $\\ 
\hline
0.1&\dmn{ -7.58}{-1}&\dmn{1.0}{-6}&\dmn{6.4}{-5}&\dmn{4.97}{-3}
&\dmn{-4.90}{-3}&\dmn{7.5}{-4}\\
\hline
0.6&\dmn{-1.35}{-1}&\dmn{-1.3}{-6}&\dmn{2.7}{-7}&\dmn{1.86}{-2}
&\dmn{-1.84}{-2}&\dmn{1.3}{-5}\\
\hline
1&\dmn{-8.94}{-2}&\dmn{-1.9}{-6}&\dmn{-2.7}{-6}&\dmn{2.34}{-2}
&\dmn{-2.32}{-2}&\dmn{2.6}{-6}\\
\hline
3&\dmn{-5.08}{-2}&\dmn{-2.3}{-6}&\dmn{-4.2}{-6}&\dmn{2.84}{-2}
&\dmn{-2.81}{-2}&\dmn{3.7}{-7}\\
\hline
\end{tabular}

\end{document}